\documentclass[useAMS,usenatbib,twocolumn]{mn2e}
\usepackage{amsmath}
\usepackage{graphicx}
\usepackage{color}
 \usepackage{multirow}

\def\simlt{\lower.5ex\hbox{$\; \buildrel < \over \sim \;$}}
\def\simgt{\lower.5ex\hbox{$\; \buildrel > \over \sim \;$}}
\def\simpt{\lower.5ex\hbox{$\; \buildrel \propto \over \sim \;$}}

\def\kms{\mbox{km s$^{-1}$}}

\def\Mpc{\mbox{Mpc}}

\definecolor{mylabelcolor}{rgb}{0.5,1,1}

\title{Neutrino Masses, Dark Energy and the Gravitational
  Lensing of Pregalactic HI}

\date{\today}

\author[R.B. Metcalf]{R. Benton Metcalf \\ Max Plank Institut f\"ur Astrophysics,  
Karl-Schwarzchild-Str. 1, 85741 Garching, Germany}

\begin{document}

\maketitle

\begin{abstract}
We study the constraints which the next generation of radio telescopes could place on the mass and number of neutrino species by studying the
gravitational lensing of high redshift 21~cm emission in combination with wide-angle surveys of galaxy lensing.  We use simple characterizations of reionization history and of proposed
telescope designs to forecast the constraints and detectability threshold for neutrinos.  It is found that the degeneracy between neutrino parameters and dark energy parameters is significantly reduced by incorporating 21~cm lensing.
The combination of galaxy and 21~cm lensing could constrain the sum of the neutrino masses to within $\sim 0.04$~eV and the number of species to within $\sim 0.1$.  This is an improvement of a factor of 2.6 in mass and 1.3 in number over a galaxy lensing survey alone. This includes marginalizing over an 11 parameter cosmological model with a two parameter model for the dark energy equation of state.  If the dark energy equation of state is held fixed at $w\equiv p/\rho=-1$ the constraints improve to $\sim 0.03$~eV and $0.04$.  These forecasted errors depend critically on the fraction of sky that can be surveyed in redshifted 21~cm emission (25\% is assumed here) and the redshift of reionization ($z=7$ is assumed here). 
It is also found that neutrinos with masses too small to be detected in the data could none the less cause a significant bias in the measured dark energy equation of state.
\end{abstract}


\section{introduction}
\label{sec:introduction}

One of the most tantalizing questions in experimental particle physics -- the nature of neutrino mass -- is connected to one of the most tantalizing questions in observational cosmology -- the nature of dark energy.    Cosmological observations presently put the most stringent upper limits on the absolute mass of neutrinos \citep{2005PhRvD..71j3515S,2007ApJS..170..377S} and might continue to do so for some time to come.  However, a partial degeneracy between neutrino masses and the dark energy's equation of state limits how well either one can be determined from cosmological measurements based on the evolution of structure formation.

Atmospheric and solar neutrino oscillation experiments strongly indicate that neutrinos 
have mass and that the sum of their masses is larger than the measured mass splittings $ \sum_\nu m_\nu \simgt \sqrt{\Delta m_{\rm atm}^2} + \sqrt{\Delta m_{\rm sun}^2} \simeq 0.05~\mbox{eV}$.  Direct measurements from $\beta$-decay experiments give an upper bound on the electron neutrino mass of $2.5$~eV, there are no competing measurements of the other flavors (see \cite{2006PrPNP..57..742F} for a review of experimental results).  Combinations of cosmological data give a constraint of $\sum_\nu m_\nu < 0.66~\mbox{eV}$ assuming a flat cosmology with a cosmological constant, but this limit is considerably looser if the density of dark energy is allowed to evolve with time \citep{2005PhRvD..71j3515S,2005PhRvL..95v1301H,2007ApJS..170..377S}.

Massive neutrinos in the appropriate mass range are relativistic when
they decouple from the photons, electrons and baryons when the temperature of the universe 
is a few MeV.  They have the effect of
suppressing the power spectrum of density fluctuations on small scales (scales smaller than the horizon
when the neutrinos become non-relativistic, $k > k_{\rm nr} \simeq 0.018
\sqrt{ m_\nu \Omega_{ m}}~h~{\rm Mpc}^{-1}~{\rm eV}^{-1/2}$ where
$k$ is the Fourier wave number, $m_\nu$ is the neutrino's mass and $h$ is the Hubble parameter in units of $100~\kms\Mpc^{-1}$) due to free-steaming.  This suppression begins
at the time of decoupling and continues until today because the
thermal velocity of the neutrinos is still significant ($v_{\rm
  therm} \simeq 150~(1+z) ~ m_\nu ~{\rm km}~{\rm s}^{-1}{\rm eV}^{-1}$
when non-relativistic).  The free-streaming scale continues to
decrease from its maximum of $k_{\rm nr}$ as $k_{\rm fs}\simeq 0.82
\frac{\sqrt{ \Omega_m (1+z)^3 + \Omega_\Lambda}}{(1+z)^2}
\left( \frac{m_\nu}{\rm eV} \right) h~{\rm Mpc}^{-1}$.  Neutrinos will
resist falling into dark matter halos with velocity dispersions $\simlt v_{\rm therm}$.  For a review of neutrinos in cosmology see \cite{2006PhR...429..307L}.

It has been proposed that neutrino masses could be measured by the
next generation of weak gravitational lensing surveys
(EUCLID\footnote{www.dune-mission.net},LSST\footnote{www.lsst.org},PanSTARRS\footnote{pan-stars.ifa.hawaii.edu}, DES\footnote{https://www.darkenergysurvey.org/})
\citep{2006JCAP...06..025H,2008PhRvD..77j3008K} and that the existence of massive neutrinos may limit the ability of
these surveys to measure properties of the dark energy \citep{2005PhRvL..95v1301H,2007arXiv0709.0253K}.  Dark
energy causes structure formation to evolve differently than it would
otherwise at low redshift ($z\simlt 1$). A degeneracy arises between this and the late-time
free-streaming of massive neutrinos since at scales significantly
smaller than $k_{\rm fs}$ the suppression is scale independent and
the lensing surveys will be sensitive to a limited range in $k$.
However, at $z>1$ dark energy is expected to have negligible effects
on structure formation so if higher redshifts can be probed the
degeneracy can be removed.  The CMB (cosmic microwave background)
provides information on the power spectrum at $z\sim 1100$, but Silk
damping limits its sensitivity to scales below $k_{\rm nr}$.
Gravitational lensing of pregalactic 21-cm radiation can provide a
probe of structure formation on the scales needed, at the redshifts needed.

The prospects for measuring gravitational lensing of the pregalactic
21~cm radiation have been studied by a number of authors \citep{ZandZ2006,metcalf&white2006,HMW07,2007arXiv0710.1108L,metcalf&white2007}.  The methods used in this paper are described in more detail in \cite{metcalf&white2007}.

\section{formalism \& modeling}

\subsection{pregalactic 21 cm radiation}

A thorough review of pregalactic 21 cm radiation can be found in
\cite{astro-ph/0608032}.  For this paper we will mention only the
barest essentials needed to specify our model for 21 cm emission.
The fluctuations in the 21~cm brightness temperature depend on the spin
temperature, $T_s$, the ionization fraction, $x_{\rm H}$ and the
density of HI through
\begin{align}
\delta T_b \simeq 24  (1+\delta_b) x_{\rm H}\left( \frac{T_s- T_{\rm
CMB}}{T_s}\right)\left(\frac{\Omega_b h^2}{0.02}\right) \nonumber \\ 
\times \left( \frac{0.15}{\Omega_m h^2}\frac{1+z}{10} \right)^{1/2} 
\mbox{ mK} 
\end{align}
\citep{1959ApJ...129..536F,1997ApJ...475..429M}.  As is commonly done,
we will assume that the spin temperature is much greater than the CMB
temperature.  This leaves fluctuations in $x_H$, and the baryon
density $\delta_b = (\rho_b - \overline{\rho}_b)/\overline{\rho}_b$ as
the sources of brightness fluctuations.  We will make the simplifying
assumption that $x_{\rm H}=1$ until the universe is very rapidly and
uniformly reionized at a redshift of $z_{\rm reion}$.  We will take
$\delta_b$ to be distributed in the same way as dark matter according
to the CDM model.  Nonlinear structure formation \citep{peac96} and
linear redshift distortion \citep{1987MNRAS.227....1K} are included.
Realistically,
the reionization process will be inhomogeneous and may extend over a
significant redshift range.  This will increase
$C_\nu(\ell)$ by perhaps a factor of 10 on scales larger
than characteristic size of the ionized bubbles
\citep{2004ApJ...608..622Z} and make the distribution non-Gaussian.

\subsection{gravitational lensing}

It is convenient to express the lensing formalism in terms of the
convergence, $\kappa(\vec{\theta},z_s)$, at a position $\vec{\theta}$
on the sky which to an excellent approximation is related directly to 
the distribution of matter through which the light passes
\begin{align}
\kappa\left(\vec{\theta},z_s\right) & = 
\frac{3}{4} H_o \Omega_m \int_0^\infty dz ~ \frac{(1+z)}{E(z)}
g\left(z,z_s \right) \delta\left(\vec{\theta},z\right) \label{eq:kappa_cont}\\
  & \simeq 
\frac{3}{4} H_o \Omega_m  \sum_i \delta\left(\vec{\theta},z_i\right) \int_{z_i-\delta
  z}^{z_i+\delta z} dz ~ \frac{(1+z)}{E(z)} g\left(z,z_s
\right)  \nonumber \\  
& =  \sum_i G(z_i,z_s) \delta\left(\vec{\theta},z_i\right) \label{eq:kappa_sum2}
\end{align}
with
\begin{equation}
g(z,z_s) = \int_{z}^\infty dz' ~ \eta\left(z',z_s\right)  
\frac{D(z,0)D(z',z)}{D(z',0)}.
\end{equation}
$H_o$ is the Hubble parameter.  The convergence can be thought of as a projected dimensionless
surface density.   The weighting function for the source distance distribution,
$\eta(z)$, is normalized to unity.  $D(z',z)$ is the angular size
distance between the two redshifts and $\delta(\vec{x},z)$ is the
fractional density fluctuation at redshift $z$ and perpendicular
position $\vec{x}$.  The function
\begin{equation}
E(z)=\sqrt{\Omega_m(1+z)^3 + \Omega_\Lambda (1+z)^{3f(z)}},
\end{equation}
where $\Omega_m$ and $\Omega_\Lambda$, are the present day densities
of matter and dark energy measured in units of the critical density.
It is assumed that the universe is flat -- $\Omega_m+\Omega_\Lambda =1$.
The function describing the evolution of dark energy with redshift can
be written
\begin{equation}
f(z)=\frac{-1}{\ln(1+z)} \int^0_{-\ln(1+z)} \hspace{-0.6cm} \left[
  1+w(a) \right]  ~ d\ln a 
\end{equation}
where $w(a)$ is the equation of state parameter for the dark energy  -- the
ratio of the of its pressure to its density -- and $a=(1+z)^{-1}$ is the
scale parameter.  

For our purposes it is convenient to express equation~(\ref{eq:kappa_sum2}) as
a matrix equation,
\begin{equation}
{\it \bf K} = {\rm \bf G} \boldsymbol{\delta},
\end{equation}
where the the components of ${\bf K}$ are the convergences running
over all position angles $\vec{\theta}$ and source redshifts, $z_s$.
The components of the vector $\boldsymbol{\delta}$ run over all
position angles and foreground redshifts $z_i$.  The matrix ${\rm \bf
G}$ is a function of most of the global cosmological parameters --
$\Omega_m$, $\Omega_\Lambda$, $w$, etc -- and is independent of
position on the sky.  This equation holds equally well if $\kappa$ and $\delta$ are expressed spherical harmonic space or in the $u$-$v$ plane where interferometer observations are carried out.  

Within frequency bands of size 1~MHz we use an estimator for the Fourier transform of the convergence, $\kappa(\ell,z_s)$, denoted
$\hat{\kappa}(\ell,z_s)$, based on the \cite{ZandZ2006} estimator and
then combine the bands as in \cite{metcalf&white2007}.  
The noise in $\hat{\kappa}(\ell,\nu)$ within one frequency band  is
\begin{align}\label{N_L_nocorr}
N^{\hat{\kappa}}(\ell,\nu) 
 = \frac{(2\pi)^2}{2}
 ~~~~~~~~~~~~~~~~~~~~~~~~~~~~~~~~~~~~~~~~~~~~~~~~~~~~~~~
 \nonumber \\ \times 
\left[\sum_k \int d^2\ell'~ \frac{ \left[ \boldsymbol{\ell}\cdot \boldsymbol{\ell}'\, C_\nu(\ell',k) + \boldsymbol{\ell}\cdot (\boldsymbol{\ell}-\boldsymbol{\ell}')\, C_\nu(|\boldsymbol{\ell}'-\boldsymbol{\ell}|,k) \right]^2 }{C^T_\nu(\ell',k) C^T_\nu(|\boldsymbol{\ell}'-\boldsymbol{\ell}|,k)} \right]^{-1}
\end{align}
where $C^T_\nu(\ell,k)$ is the power spectrum of the actual brightness
temperature, while $C_\nu(\ell,k)=C^T_\nu(\ell,k) +
C^N_\nu(\ell,k) $ is the observed power spectrum which
includes noise.  $\ell$ is the angular Fourier mode number. Because
the estimator is a sum over all the 
observed pairs of visibilities, it will (by the central limit theorem)
be close to Gaussian distributed even though it is quadratic in the
visibilities. 

The $\hat{\kappa}(\ell,z)$s can be grouped into a data vector 
\begin{align}
  \label{eq:datavector}
  {\bf D} &= \hat{\bf K}-{\bf K} \\
  &= \hat{\bf K}-{\bf G} \boldsymbol{\delta}
\end{align}
where the components run over all the combinations of $z_i$ and $\ell$
that are measured.  To a good approximation modes from different bands and $\ell$'s separated by more than the resolution of the telescope are statistically independent so the covariance matrix for ${\bf D}$ is diagonal
\begin{align}
 {\rm\bf N}_{ij} & \simeq  \delta_{ij} N^{\hat{\kappa}}(\ell_i,\nu_i). \label{eq:diagonal-noise2}
\end{align}
with $|\ell_i-\ell_j|$ larger than the resolution.
The likelihood function for the correlations in
$\kappa(\ell,z_i)$ is given by
\begin{equation}
  \label{eq:likelihood2}
 \ln {\mathcal L} = - \frac{1}{2}  \hat{\bf K}^\dag {\rm\bf C}^{-1} \hat{\bf
     K}  - \frac{1}{2} |{\rm\bf C}|, 
\end{equation}
where 
\begin{equation}
{\bf C} = {\bf N} + {\bf C}_\kappa
\end{equation}
\citep{metcalf&white2007}.   ${\bf C}_\kappa$ here is the
(cross-)power spectrum of the convergence for two different source redshifts,
\begin{equation}
\left[{\bf C}_\kappa\right]_{ij} = \left\langle
\kappa(\vec{\ell},z_i) \kappa(\vec{\ell},z_j) \right\rangle.
\end{equation}
This can be calculated using expression~(\ref{eq:kappa_cont}) and a
model for the matter power spectrum.

\subsection{structure formation with massive neutrinos}

To calculate the linear matter power spectrum we use the analytic
formulae of \cite{1999ApJ...511....5E} with the modification of
\cite{2007arXiv0709.0253K} that improves accuracy when the neutrino
masses are small.  We use the method of
\cite{peac96} to transform this linear power spectrum into a power spectrum
with nonlinear structure formation.  This method has not been tested thoroughly against simulations of
nonlinear structure formation with massive neutrinos, but we
expect that the relation between the linear and nonlinear power
spectrum will be only weakly affected by this.  There is no baryon acoustic oscillations in the power spectrum.

We assume that there are $N_\nu$ species of neutrinos with the same
mass.  The total density of neutrinos is fixed by the physics at the
time of decoupling to be $\Omega_\nu h^2 = \sum_\nu m_\nu/93.14~
{\rm eV}$.  If the masses are not degenerate the
measured $N_\nu$ will not be an integer.  These are the parameters we
will try to constrain.  Roughly speaking $N_\nu$ controls the average mass of 
the neutrino species and through this the free-streaming scale while 
$\Omega_\nu h^2$ affects the degree of suppression to the power spectrum.

The measurements we consider
are actually sensitive to any light particle species with significant
cosmological densities not just neutrinos.  The interpretation of the
parameters would be different for relic particles that were produced
in a different way, for example axions, but the late-time physics
would be the same.

\subsection{model observations}

\subsubsection{21 cm observations}

We will concentrate on the planned SKA (the Square Kilometer
Array)\footnote{www.skatelescope.org/} because it is the only planned
telescope that will have a large enough size to be relevant to
neutrino constraints.  It is only the core of the telescope that will
be used for observing pregalactic 21~cm radiation.
Plans for the SKA core have not been finalized, but it
is expected to have a diameter of
$D_{\rm tel} \sim 6$~km ($\ell_{\rm max} \sim 10^4$),
an aperture covering fraction of $f_{\rm cover} \sim 0.02$ (the total
collecting area of the telescopes divided by $\pi (D_{\rm tel}/2)^2$) and a frequency range
extending down to $\sim 100$~MHz which corresponds to $z\sim 13$.  It is expected that
the SKA will be able to map the 21~cm emission with a resolution of
$\Delta\theta \sim 1\mbox{ arcmin}$. For reference, one arcminute
(fwhm) corresponds to baselines of 5.8~km at $z=7$ and 11~km at
$z=15$.  We will take $\ell_{\rm min}=10$ to be the lowest mode to be
measured. 

The noise in each visibility measurement will have a thermal component
and a component resulting from imperfect foreground subtraction.  Here
we model only the thermal component.  If the telescopes in the array
are uniformly distributed on the ground, the average integration time
for each baseline will be the same and the power spectrum of the noise
will be
\begin{align}\label{eq:C_noise}
 C_\ell^N = \frac{2\pi}{ \Delta\nu t_o} \left(  \frac{T_{\rm sys} \lambda}{ f_{\rm cover} D_{\rm tel}} \right)^2 = \frac{(2\pi)^3 T_{\rm sys}^2}{\Delta\nu t_o f_{\rm cover}^2 \ell_{\rm max}(\nu)^2}, 
\end{align}
\citep{2004ApJ...608..622Z,2005ApJ...619..678M,2006ApJ...653..815M}
where $T_{\rm sys}$ is the system temperature, $\Delta\nu$ is the
bandwidth, $t_o$ is the total observation time, and $\ell_{\rm max}(\lambda)=2\pi D_{\rm
tel}/\lambda$ is the highest multipole that can be measured by the
array, as set by the largest baselines.  
At the relevant frequencies, the overall system temperature is
expected to be dominated by galactic synchrotron radiation.  We will
approximate the brightness temperature of this foreground as $T_{\rm
sky}=180 \mbox{ K}(\nu/180\mbox{ MHz})^{-2.6}$, as appropriate for
regions well away from the Galactic Plane \citep{astro-ph/0608032}.
This results in larger effective noise for higher redshift
measurements of the 21~cm emission.  We will consider an observation
time of 90~days which might be achievable within three seasons of
observation and we will assume that the survey covers 25\% of the sky.

\subsubsection{galaxy weak lensing survey}

For comparison and for combining with the 21~cm lensing we will
consider a model galaxy lensing survey.
The noise in power spectrum estimates from such a
survey can be written as $N_\kappa(\ell) = \sigma^2_\epsilon / n_{\rm
g}$ where $n_{\rm g}$ is the angular number density of background
galaxies and $\sigma_\epsilon$ is the root-mean-square intrinsic
ellipticity of those galaxies.  This neglects all systematic errors as
well as photometric redshift uncertainties.  Following standard
assumptions, we model the redshift distribution of usable galaxies as
$\eta(z) \propto z^2 e^{-(z/z_o)^{1.5}}$, where $z_o$ is set by the
desired median redshift, and we adopt $\sigma_\epsilon = 0.25$.  The
EUCLID satellite (the imaging part of which was previously known as
DUNE\footnote{www.dune-mission.net} ) proposes to survey 20,000 square
degrees on the sky to a usable galaxy density of $n_{\rm g}\simeq
35\mbox{ arcmin}^{-2}$ with a median redshift of $z\sim 0.9$. Several
planned ground-based surveys -- LSST,
PanSTARRS --  will cover comparable areas to EUCLID at
a similar depth.  In order to use tomographic information, we
divide the galaxies into 10 redshift bins each containing the same
number of galaxies.  

\subsubsection{CMB observations}

The Planck
Surveyor\footnote{www.rssd.esa.int/index.php?project=Planck} will
do a full sky survey of the CMB radiation with higher resolution and
more sensitivity to polarization than is now available.  We incorporate
these future measurements into our forecasts by calculating the
expected Fisher matrix.  To do this we use the same technique as
described in the appendix of \cite{2008arXiv0810.0003R}.  This includes the contributions from both the scalar and tensor perturbations.

\subsection{the cosmology}

We use a 11 parameter cosmological model.  The energy densities in dark
energy and baryons are $\Omega_\Lambda$ and $\Omega_b$ in units of the critical density.  The density of matter is fixed to $\Omega_{ m}=1-\Omega_\Lambda$ to make the
geometry flat.    The primordial power spectrum is
\begin{equation}
P_o(k) = A_s~ \left( \frac{k}{H_oc^{-1}}\right)^{n_s + \frac{dn_s}{d\ln k} \ln k}
\end{equation}
The $\frac{dn_s}{d\ln k}$ parameter is included because if the
primordial power spectrum is not a pure power-law it might partially
mock the effect of early neutrino free-streaming on the power
spectrum.  The time dependent dark energy equation of state has two parameters
$w(a)=w_o+ w_a(1-a)$.  There are two neutrino parameter
$N_\nu$ and $\sum_\nu m_\nu$.  Including $\frac{dn_s}{d\ln k}$ and a
non-constant $w$ makes this a more general set of parameters with more
potential degeneracies than has usually been used when predicting
constraints on massive neutrinos.  The optical depth to CMB last scattering surface is $\tau$.  The fiducial model is set to $\{\tau, h,\Omega_{\Lambda} h^2,\Omega_b h^2,n_s,\frac{dn_s}{d\ln k},w_o,w_a,N_\nu\}=\{ 0.09, 0.7,0.343,0.0223,1,0,-1,0,3\}$.  The normalization, $A_s$, is set so that the fluctuations within a sphere of radius 8~Mpc is $\sigma_8=0.75$ in the fiducial model.  Two values for $\Sigma_\nu m_\nu$ are used, 0.66~eV and 0.09~eV.  All calculated parameter constraints are marginalized over the other parameters.

\section{results}
\label{sec:results}

To assess how well the proposed observations could measure neutrino
properties we adopt two statistical methods that are popular in the
literature -- Fisher matrix forecasts and the Bayesian evidence method.  

\subsection{Fisher matrix forecast}
\label{sec:fish-matr-forec}

The maximum likelihood estimate for any parameter can be found by
maximizing~(\ref{eq:likelihood2}) with respect to that parameter.  The
error in this estimator is often forecast using the Fisher matrix
defined as
\begin{equation}\label{eq:likely_diag}
 {\rm\bf F}_{ij} = - \left\langle \frac{\partial^2
     \ln\mathcal{L}}{\partial p_i \partial p_j} \right\rangle.
\end{equation} 
The expected error in the parameter $p_a$, marginalized over all other
parameters, is $\sigma_a^2 \simeq \left({\rm\bf
F}^{-1}\right)_{aa}$. The unmarginalized error estimate (the error
when all other parameters are held fixed) is $\left( {\rm\bf
F}_{aa}\right)^{-1}$.

Since $\ell$-modes separated by more than the resolution of the
telescope will not be correlated, we can break the likelihood function
up into factors representing each resolved region in $\ell$-space
\citep{metcalf&white2006}.  The result is that there are $\sim
(2\ell+1)f_{\rm sky}$ independent measured modes for each value of
$\ell$, where $f_{\rm sky}$ is the fraction of sky surveyed.  The
Fisher matrix can then be further simplified to the widely used form,
\begin{equation}\label{eq:fisher}
{\bf F}_{ab} = \frac{1}{2} \sum^{\ell_{\rm max}}_{\ell=\ell_{\rm min}} (2\ell +1) f_{\rm sky} {\rm tr}\left[ {\bf C}^{-1} {\bf C},_a {\bf C}^{-1} {\bf C},_b \right]~.
\end{equation}

\begin{figure} \rotatebox{90}{
\includegraphics[width=6.0cm]{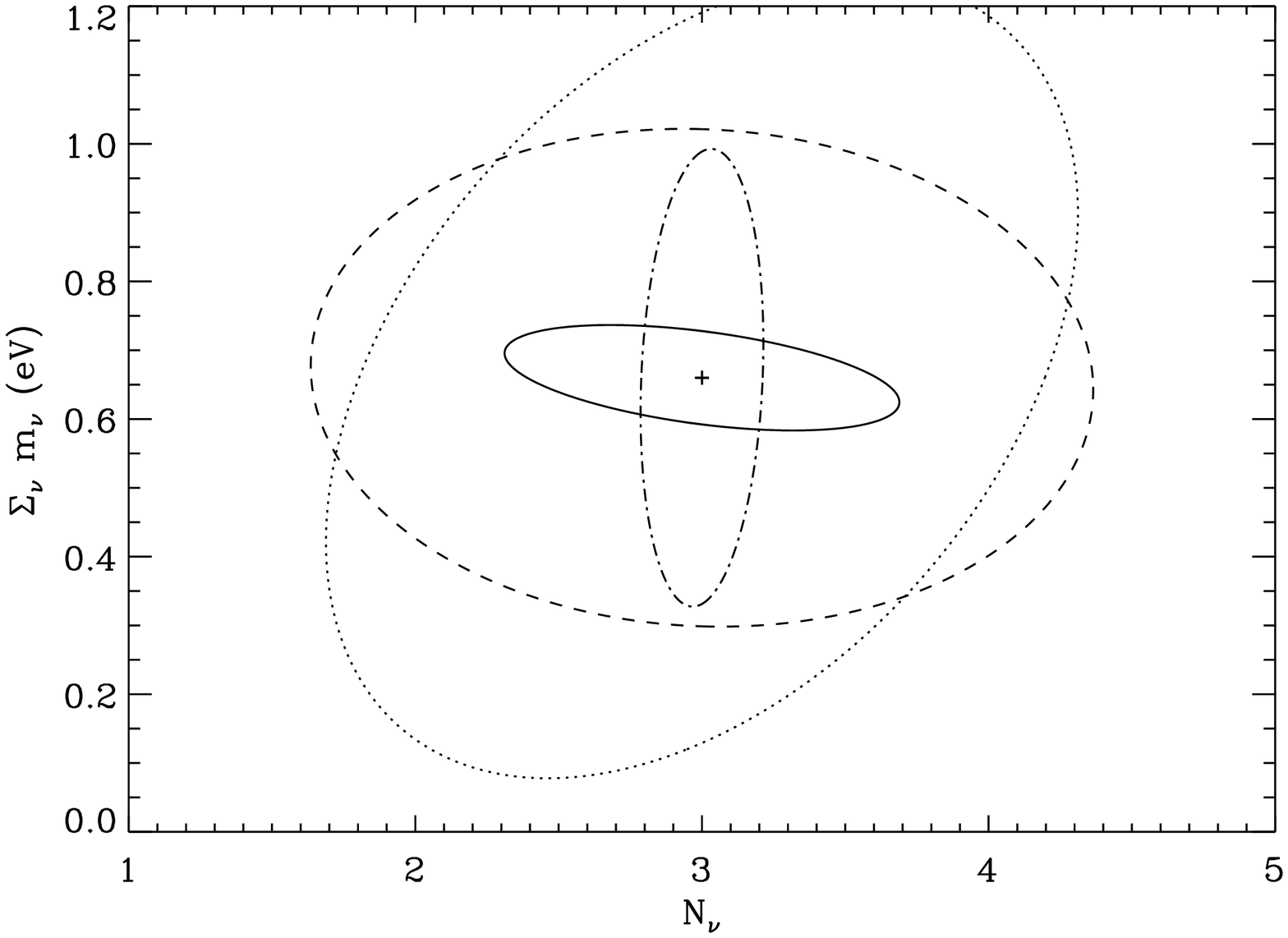}} 
\caption{The forecasted constraints on the combined mass and number of neutrinos with the fiducial model $\{\sum_\nu m_\nu,N_\nu\}=\{0.66~\mbox{eV},3\}$ marginalized over all the other cosmological parameters.  The dotted curves are for a EUCLID-like galaxy lensing survey, the dashed curves are for a 21~cm lensing survey, the solid curves are for the tomographic combination of the two surveys and the dot-dashed curve is for Planck by itself.} 
\label{fig:neutrinos_dn}
\end{figure}

\begin{figure} \rotatebox{90}{
\includegraphics[width=6.0cm]{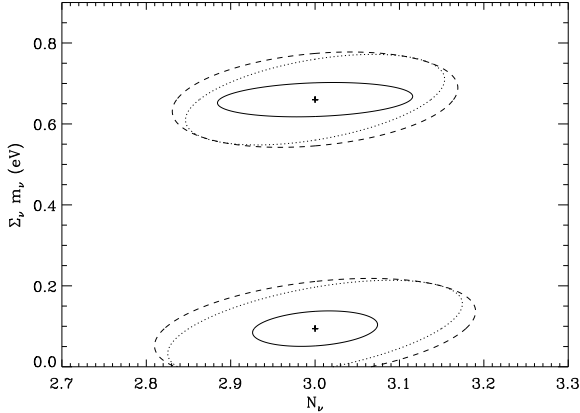}} 
\caption{The same as figure~\ref{fig:neutrinos_dn} except the Planck constraints are combined with the lensing constraints and the scale has been changed.  There are two fiducial models at  $\{\sum_\nu m_\nu,N_\nu\}=\{0.66~\mbox{eV},3\}$ and  $\{\sum_\nu m_\nu,N_\nu\}=\{0.09~\mbox{eV},3\}$. } 
\label{fig:neutrinos_planck_dn}
\end{figure}

\begin{figure} \rotatebox{90}{
\includegraphics[width=6.0cm]{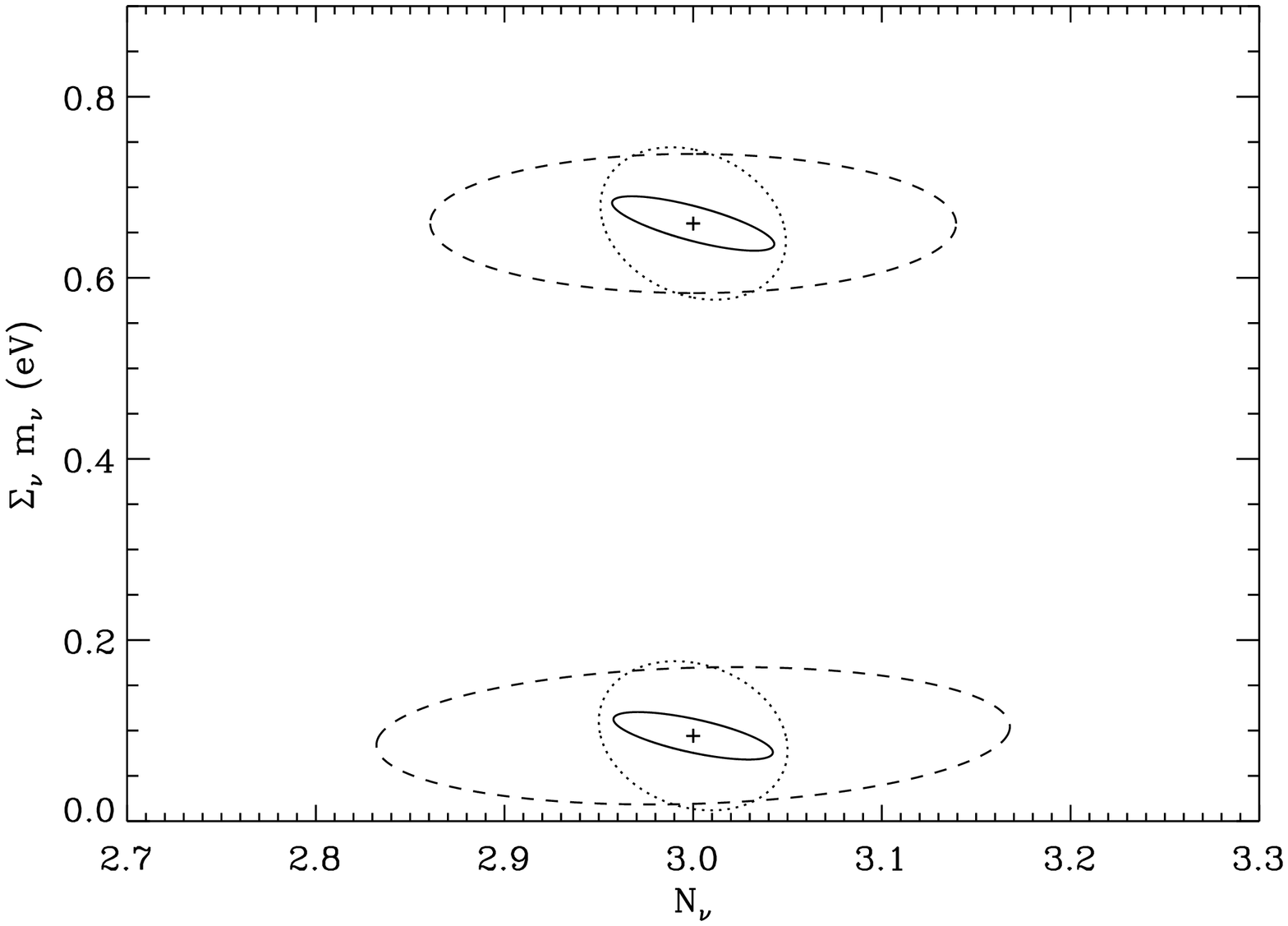}} 
\caption{Same as figure~\ref{fig:neutrinos_planck_dn} except the dark matter equation of state is held fixed at $\{ w_o,w_a \}=\{-1,0\}$. } 
\label{fig:neutrinos_planck_fixw_dn}
\end{figure}

\begin{figure} \rotatebox{90}{
\includegraphics[width=6.0cm]{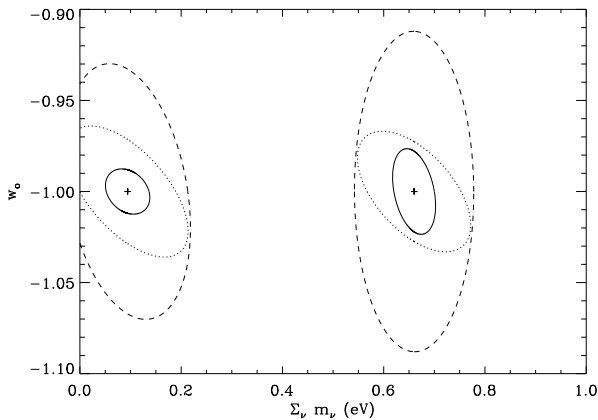}} 
\caption{The constraints in the $w_o$ -- $\sum_\nu m_\nu$ plane.  Planck constraints are included.  Note the improvement in the $w_a$ constraints especially when the neutrino masses are small.} 
\label{fig:neutrino_wvsm_bw.ps}
\end{figure}

The error estimates on the neutrino parameters are shown in
figures~\ref{fig:neutrinos_dn} through
\ref{fig:neutrinos_planck_fixw_dn} for combinations of different data
and model assumptions.  Figure~\ref{fig:neutrinos_dn} shows the
constraints using the three kinds of data by themselves and the
combination of galaxy and 21~cm lensing.  These lensing surveys are
combined tomographically not just by adding their Fisher matrices as
would be the case if they were independent measurements.  Ten redshift bins 
with equal numbers of galaxies are used and ten redshift bins of fixed z-width 
are used for the 21~cm lensing.  The CMB does a relatively good job of 
constraining the $N_\nu$, but not as good a job of constraining the sum of the masses.

In figure~\ref{fig:neutrinos_planck_dn} the CMB constraints are
combined with the lensing constraints.  There is a drastic improvement
because the parameter degeneracies are significantly reduced (Note the change in scale from figure~\ref{fig:neutrinos_dn}).  The
forecasted errors are 0.04~eV and 0.1 for the combination of lensing
surveys with $\sum_\nu m_\nu = 0.66$.  This is an improvement of 2.6 in mass and 1.3 in $N_\nu$ over either of the lensing
surveys by themselves.   In figure~\ref{fig:neutrinos_planck_fixw_dn}
the dark energy equation of state is kept fixed ($w=-1$).  The
improvement in the constraints illustrates the degeneracy between
neutrinos and dark energy parameters.   The effect on the galaxy
lensing constraints are particularly strong.  If one only considers 
cosmological constant models the constraints improve to 0.03~eV and 0.04 for the
combined lensing case.

Figure~\ref{fig:neutrino_wvsm_bw.ps} shows the constraints in the $w_o$ - $\sum_\nu m_\nu$ plane.  It can be seen more directly here how the degeneracy between dark energy and massive neutrinos is reduced by including 21~cm lensing, especially when the neutrino masses are small.  Note also that if it were assumed that neutrinos are massless, the maximum likelihood would not give the correct value for $w_o$.

These calculations are in agreement with previous calculations of the constraints from future galaxy lensing surveys \citep{2008PhRvD..77j3008K,2006JCAP...06..025H} where more restricted cosmological models were used and thus stronger constraints were found.

\subsection{Bayesian evidence}

Another statistical question that could be asked is whether the data
{\it requires} massive neutrinos.  One way to answer this question is
using the average ratio of the Bayesian evidence \citep{jeffreys1961} for models with and without massive neutrinos.  This technique has been used extensively in cosmological parameter estimation
(\cite{2006PhRvD..74l3506L,2004MNRAS.348..603S,2007MNRAS.380.1029H} for example) and galaxy lensing surveys specifically \citep{2008PhRvD..77j3008K}.

The Bayesian evidence is the probability of the data, $D$, given the
model, $M$, marginalized over the parameters of that model, $\{\theta\}$,
\begin{align}
E(D|M)= \int d\theta~ p(D|\theta,M)p(\theta|M).
\end{align}
The integral is over all of the parameter space.
Bayes factor is the ratio of the evidences for two competing models
\begin{align}
B  \equiv  \frac{E(D|M_o)}{E(D|M_1)}.
\end{align}
Here model $M_o$ is the simpler model with $n_o$ parameters and model $M_1$
is more complex and has more parameters, $n_1>n_o$.

To make a forecast, Bayes factor must be averaged over expected data
sets.  It will be assumed that model $M_o$ is the real case so that
the averaging is according to this model.  We are asking how well
we can expect to role out model $M_1$.  In calculating this we use the
approximation to $\langle B \rangle$ derived by \cite{2007MNRAS.380.1029H} the Savage-Dickey ratio
\begin{align}
\langle B \rangle  \simeq  (2\pi)^{-(n_1-n_o)/2} \sqrt{\frac{ |F^{(1)}| }{ |F^{(o)}|}} \exp\left[-\frac{1}{2}
  \delta\theta_{\nu} F^{(1)}_{\nu\mu} \delta\theta_\mu\right] 
\end{align}
where 
\begin{align}
\delta\theta_i = \left[ \left( F^{(o)} \right)^{-1} \right]_{ij}F^{(1)}_{j\alpha}\delta\theta_\alpha 
\end{align}
and the range of the indexes are $i=1...n_o$, $\alpha=n_o+1 ... n_1$ and $\mu,\nu = 1 ... n_1$.  We have assumed that there are no a priori limits on the range of the neutrino parameters.

\begin{figure} \rotatebox{90}{
\includegraphics[width=6.0cm]{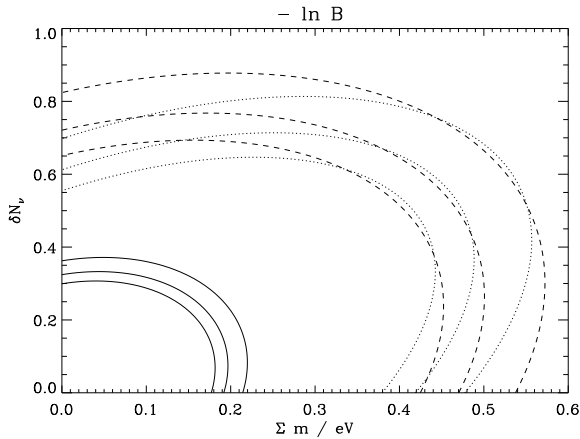}} 
\caption{The log of the average evidence ratio between a model with and without massive neutrinos.  The axis are the deviation from the fiducial model $\{\sum_\nu m_\nu,N_\nu\} = \{0,3\}$.  The three contour curves of each type are for 1, 2.5 and 5.  According to 
Jeffreys (1961)
$|\ln B| < 1$ is ``inconclusive'', $1< |\ln B| < 2.5$ is
``substantial'', $2.5< |\ln B| < 5$ is ``strong'' and $|\ln B | > 5$ is ``decisive'' evidence for the additional parameters. The expected constraints from Planck have been incorporated.  The dotted curves are for a EUCLID-like galaxy lensing survey, the dashed curves are for a 21~cm lensing survey and the solid curves are for the tomographic combination of the two surveys.} 
\label{fig:neutrinos_evidence}
\end{figure}

\begin{figure} \rotatebox{90}{
\includegraphics[width=6.0cm]{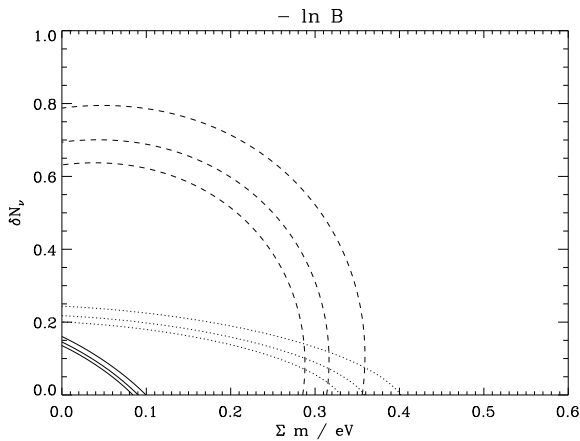}} 
\caption{Same as figure~\ref{fig:neutrinos_evidence} except the dark matter equation of state is held fixed at $\{ w_o,w_a \}=\{-1,0\}$.} 
\label{fig:neutrinos_evidence_fixw}
\end{figure}

Figure~\ref{fig:neutrinos_evidence} shows evidence ratio as a function of the deviation from the fiducial model, $\{\sum_\nu m_\nu,N_\nu\} = \{0,3\}$.  Again we see that the constraints from the lensing of galaxies and 21~cm put similar constraints on the neutrino parameters.  By combining these data sets the constraints are improved by more than a factor of two in each dimension.  A combined mass as small as $0.2$~eV should be detected, $0.19$~eV if $N_\nu$ is fixed to 3.   
Figure~\ref{fig:neutrinos_evidence_fixw} shows how these constraints change if the dark energy equation of state is fixed to the cosmological constant value, $\{w_o,w_a\}=\{-1,0\}$.  The constraints drastically improve which again demonstrates the degeneracy between dark energy and massive neutrinos.  In this case, a combined mass of $0.09$~eV would be detectable.

\section{Discussion}

If it is assumed that neutrinos are massless, or have too small a mass to be of significance, the 
dark energy equation of state, $w=p/\rho$, could be underestimated from cosmological constraints.  
This would be the case even if the data does not allow for a detection of neutrino mass.  It is also true 
that any independent constraint on the neutrino mass would improve future cosmological constraints on 
dark energy.  The KATRIN\footnote{www-ik.fzk.de/katrin/} beta-decay experiment expects to reach a level 
of $0.2$~eV for the electron neutrino mass and thus might have an impact on dark energy constraints.  In section~\ref{sec:results} it was shown that a neutrino mass as small as $\sum_\nu m_\nu /N_\nu = 0.03$~eV could 
bias $w_o$ high by $\sim 1~\sigma$ in future galaxy lensing surveys if it is not accounted for and if it is accounted for the error bars increase by a factor of 3.4 for $w_o$ and 2 for $w_a$.  Lensing of high redshift 21~cm is a way to reduce this 
degeneracy by adding constraints on the growth of structure at higher redshifts where dark energy is 
presumed to contribute very little.  Type Ia Supernovae surveys 
and surveys aimed at measuring the baryon acoustic oscillations at $z \sim 1$ will put constraints on 
dark energy that are based on the luminosity distance redshift relation and so are independent of the growth of structure.  It is important that methods based both on structure formation and on luminosity distance are 
fully realized to avoid systematic errors.  The two probes are also necessary to test alternative theories of 
gravity on large scales which might also provide an explanation for the apparent acceleration of the cosmological expansion.

\vspace{0.3cm} 
\leftline{\bf Acknowledgments} 
I would like to thank S. White for helpful discussion and criticism.  I would also like to thank J. Weller for loaning me his code for calculating the Planck priors on the cosmological parameters.

 \bibliographystyle{/Users/bmetcalf/Work/TeX/apj/apj}
 \bibliography{/Users/bmetcalf/Work/mybib}

\end{document}